\documentclass[%
 reprint,%
 amssymb, amsmath,%
 aip,cha,%
]{revtex4-1}

\usepackage{docs}%
\usepackage{bm}%
\usepackage[colorlinks=true,linkcolor=blue]{hyperref}%
\expandafter\ifx\csname package@font\endcsname\relax\else
 \expandafter\expandafter
 \expandafter\usepackage
 \expandafter\expandafter
 \expandafter{\csname package@font\endcsname}%
\fi
\begin{document}

\title{Topological and Hopf charges of a twisted Skyrmion string}%

\author{Malcolm Anderson$^1$,~Miftachul Hadi$^{1,2}$,~Andri Husein$^3$}%
\email{itpm.id@gmail.com (Miftachul Hadi)}
\affiliation{$^1$Department of Mathematics, Universiti Brunei Darussalam, Negara Brunei Darussalam\\
             $^2$Physics Research Centre, Indonesian Insitute of Sciences, Puspiptek, Serpong, Indonesia\\
		     $^3$Department of Physics, University of Sebelas Maret, Surakarta, Indonesia}%

\begin{abstract}
We examine nonlinear sigma models, in particular the Skyrme model with a twist (the twisted Skyrmion string), which comprises a vortex solution with an added dependence on a twist term $mkz$, where $z$ is the vertical coordinate. The topological and Hopf charges of a twisted Skyrmion string are calculated and are shown to be equivalent to the winding number $n$ of the vortex and proportional to $nmk$, respectively. The principle of conservation of the Hopf charge, which is a standard property of the baby Skyrmion model, is generalised to cover the non-compact twisting solutions studied here.
\end{abstract}

\maketitle

\section{Introduction to Nonlinear Sigma Model}
A nonlinear sigma model is an $N$-component scalar field theory in which the fields are functions defining a mapping from the space-time to a target manifold \cite{zakr}. 
By a nonlinear sigma model, we mean a field theory with the following properties \cite{hans02}: 
\begin{itemize}
\item[(1)] The fields, $\phi(x)$, of the model are subject to nonlinear constraints at all points $x\in\mathcal{M}_0$, where $\mathcal{M}_0$ is the source (base) manifold, i.e. a spatial submanifold of the (2+1) or (3+1)-dimensional space-time manifold.
\item[(2)] The constraints and the Lagrangian density are invariant under the action of a global (space-independent) symmetry group, $G$, on $\phi(x)$.
\end{itemize}

The Lagrangian density of a free (without potential) nonlinear sigma model on a Minkowski background space-time is defined to be \cite{chen}
\begin{equation}\label{1}
\mathcal{L}=-\frac{1}{2\lambda^2}~\gamma_{AB}(\phi)~\eta^{\mu\nu}~\partial_\mu\phi^A~\partial_\nu\phi^B
\end{equation}
where $\gamma_{AB}(\phi)$ is the field metric, $\eta^{\mu\nu}=\text{diag}(1,-1,-1,-1)$ is the Minkowski tensor, $\lambda$ is a scaling constant with dimensions of (length/energy)$^{1/2}$ and $\phi={\phi^A}$ is the collection of fields. Greek indices run from 0 to $d-1$, where $d$ is the dimension of the space-time, and upper-case Latin indices run from 1 to $N$.  

The simplest example of a nonlinear sigma model is the $O(N)$ model, which consists of $N$ real scalar fields, $\left\{\phi\right\}_{A=1}^N$, with the Lagrangian density \cite{hans02}
\begin{equation}\label{2}
\mathcal{L}=-\frac{1}{2\lambda^2}~\delta_{AB}~\eta^{\mu\nu}~\partial_\mu\phi^A~\partial_\nu\phi^B
\end{equation}
where the scalar fields, $\phi^A$, satisfy the constraint
\begin{equation}\label{3}
\delta_{AB}~\phi^A\phi^B=1
\end{equation}
and $\delta_{AB}$ is the Kronecker delta. The Lagrangian density (\ref{2}) is obviously invariant under the global (space-independent) orthogonal transformations $O(N)$, i.e. the group of $N$-dimensional rotations \cite{hans02}
\begin{equation}\label{4}
\phi^A\rightarrow\phi'^A=O^A_B~\phi^B.
\end{equation}

One of the most interesting examples of a $O(N)$ nonlinear sigma model, due to its topological properties, is the $O(3)$ nonlinear sigma model in 1+1 dimensions, with the Lagrangian density \cite{hadi2014}
\begin{equation}\label{5}
\mathcal{L}_1=-\frac{1}{2\lambda^2}~\eta^{\mu\nu}~\partial_\mu\phi~.~\partial_\nu\phi 
\end{equation}
where, if $d$ is the number of space-time dimensions, the indices $\mu$ and $\nu$ run from 0 to $d-1$, and $\phi=(\phi^1,\phi^2,\phi^3)$  is subject to the constraint $\phi.\phi=1$, where the dot (.) denotes the standard inner product on the 3-dimensional real coordinate space $\mathbb{R}^3$. The target manifold is the therefore the unit sphere $\textbf{S}^2$ in $\mathbb{R}^3$.

A simple representation of $\phi$ (in the general time-dependent case) is
\begin{equation}\label{6}
\phi=
\begin{pmatrix}
\sin f(t,{\bf r})~\sin g(t,{\bf r}) \\
\sin f(t,{\bf r})~\cos g(t,{\bf r}) \\
\cos f(t,{\bf r})
\end{pmatrix}
\end{equation}
where $f$ and $g$ are scalar functions on the background space-time, with Minkowski coordinates $x^\mu=(t,{\bf r})$. In what follows, the space-time dimension, $d$, is taken to be 4, and so $\bf r$ is a 3-vector.

If we substitute (\ref{6}) into (\ref{5}), the Lagrangian density becomes
\begin{equation}\label{7}
\mathcal{L}_1=-\frac{1}{2\lambda^2}[\eta^{\mu\nu}~\partial_\mu f~\partial_\nu f+(\sin^2f)~\eta^{\mu\nu}~\partial_\mu g~\partial_\nu g]
\end{equation}
The Euler-Lagrange equations associated with $\mathcal{L}_1$ in (\ref{7}) are
\begin{eqnarray}\label{8}
\eta^{\mu\nu}~\partial_\mu\partial_\nu f-(\sin f~\cos f)~\eta^{\mu\nu}~\partial_\mu g~\partial_\nu g=0
\end{eqnarray}
and
\begin{eqnarray}\label{9}
\eta^{\mu\nu}~\partial_\mu\partial_\nu g+2(\cot f)~\eta^{\mu\nu}~\partial_\mu f~\partial_\nu g=0.
\end{eqnarray}

\section{Vortex Solution}
A simple solution to the $O(3)$ field equations (\ref{8}) and (\ref{9}) is the vortex solution found by imposing the 2-dimensional "hedgehog" ansatz, i.e.
\begin{equation}\label{10}
\phi=
\begin{pmatrix}
\sin f(r)~\sin (n\theta-\chi)\\
\sin f(r)~\cos (n\theta-\chi)\\
\cos f(r)
\end{pmatrix}
\end{equation}
where $r=(x^2+y^2)^{1/2}$, $\theta=\arctan (y/x)$, $n$ is a positive integer, and $\chi$ is a constant phase factor.

A vortex is a stable time-independent solution to a set of classical field equations that has finite energy in two spatial dimensions; it is a two-dimensional soliton. In three spatial dimensions, a vortex becomes a string, a classical solution with finite energy per unit length \cite{preskill}. Solutions with finite energy, satisfying the appropriate boundary conditions, are candidate soliton solutions \cite{manton}. The boundary conditions that are normally imposed on the vortex solution (\ref{10}) are $f(0)=\pi$ and $\lim_{r\to\infty}f(r)=0$, so that the vortex ''unwinds'' from $\phi=-\hat{\textbf{z}}$ to $\phi=\hat{\textbf{z}}$ as $r$ increases from 0 to $\infty$. 

The function $f$ in this case satisfies the field equation 
\begin{equation}\label{11}
r~\frac{d^2f}{dr^2}+\frac{df}{dr}-\frac{n^2}{r}~\sin f~\cos f=0
\end{equation}
There is in fact a family of solutions to this equation (\ref{11}) satisfying the standard boundary conditions, i.e.
\begin{equation}\label{12}
\sin f=\frac{2K^{1/2}r^n}{Kr^{2n}+1},~\text{or equivalently}~
\cos f=\frac{Kr^{2n}-1}{Kr^{2n}+1}
\end{equation}
where $K$ is positive constant.
The energy density $\sigma$ of a static (time-independent) field with Lagrangian density $\mathcal{L}_1$ in (\ref{7}) is
\begin{eqnarray}\label{13}
\sigma 
&=& -\mathcal{L}_1 \nonumber\\
&=& \frac{1}{2\lambda^2}\left[\eta^{\mu\nu}~\partial_\mu f~\partial_\nu f+(\sin^2 f)~\eta^{\mu\nu}~\partial_\mu g~\partial_\nu g\right]
\end{eqnarray}
The energy density of the vortex solution is
\begin{eqnarray}\label{14}
\sigma =\frac{4Kn^2}{\lambda^2}\frac{r^{2n-2}}{(Kr^{2n}+1)^2}
\end{eqnarray}
In particular, the total energy
\begin{equation}\label{15}
E=\int\int\int \sigma~ dx~dy~dz,
\end{equation}
of the vortex solution is infinite. But the energy per unit length of the vortex solution
\begin{eqnarray}\label{16}
\mu 
&=& \int\int \sigma~dx~dy\nonumber\\
&=& 2\pi\int_0^\infty\frac{4Kn^2}{\lambda^2}\frac{r^{2n-2}}{(Kr^{2n}+1)^2}~r~dr  = \frac{4\pi n}{\lambda^2}
\end{eqnarray}
is finite, and does not depend on the value of $K$. (We use the same symbol for the energy per unit length and the mass per unit length, due to the equivalence of energy and mass embodied in the relation $E=mc^2$. Here, we choose units in which $c=1$). 

This last fact means that the vortex solution in the nonlinear sigma model have no preferred scale. A small value of $K$ corresponds to a more extended vortex solution, and a larger value of $K$ corresponds to a more compact vortex solution, as can be seen by plotting $f$ (or $-\mathcal{L}_1$) for different values of $K$ and a fixed value of $n$. This means that the vortex solutions are what is called neutrally stable to changes in scale. As $K$ changes, the scale of the vortex changes, but the mass per unit length, $\mu$, does not. Note that because of equation (\ref{16}), there is a preferred winding number, $n=1$, corresponding to the smallest possible positive value of $\mu$.

\section{Skyrmion Vortex without a Twist} 
If a Skyrme term is added to the original sigma model Lagrangian density $\mathcal{L}_1$  (with the unit sphere as the target manifold) in (\ref{5}), the result is a modified Lagrangian density of the form
\begin{eqnarray}\label{17}
\mathcal{L}_2
&=&-\frac{1}{2\lambda^2}~\eta^{\mu\nu}~\partial_\mu\phi~.~\partial_\nu\phi \nonumber\\
&& -K_s~\eta^{\kappa\lambda}~\eta^{\mu\nu}(\partial_\kappa\phi\times\partial_\mu\phi)~.~(\partial_\lambda\phi\times\partial_\nu\phi) 
\end{eqnarray}
where the Skyrme term is the second term on the right hand side of (\ref{17}) and $K_s$ is a positive coupling constant.  

With the choice of field representation (\ref{6}), equation (\ref{17}) becomes
\begin{eqnarray}\label{18}
\mathcal{L}_2
&=&-\frac{1}{2\lambda^2}\left(\eta^{\mu\nu}~\partial_\mu f~\partial_\nu f+\sin^2f~\eta^{\mu\nu}~\partial_\mu g~\partial_\nu g\right) \nonumber\\
&&-2K_s~\sin^2f\left[\left(\eta^{\mu\nu}~\partial_\mu f~\partial_\nu f\right)\left(\eta^{\kappa\lambda}~\partial_\kappa g~\partial_\lambda g\right) \right.\nonumber\\
&&\left. -\left(\eta^{\mu\nu}~\partial_\mu f~\partial_\nu g\right)^2\right]
\end{eqnarray}
If the vortex configuration (\ref{10}) for $\phi$ is assumed, the Lagrangian density becomes
\begin{eqnarray}\label{19}
\mathcal{L}_2
&=& -\frac{1}{2\lambda^2}\left[\left(\frac{df}{dr}\right)^2 +\frac{n^2}{r^2}\sin^2f\right] \nonumber\\
&& -K_s\frac{n^2}{r^2}\sin^2f\left(\frac{df}{dr}\right)^2
\end{eqnarray}
The Euler-Lagrange equations generated by $\mathcal{L}_2$, namely
\begin{eqnarray}\label{20}
\partial_\mu\left[\frac{\partial\mathcal{L}_2}{\partial(\partial_\mu f)}\right]  -\frac{\partial\mathcal{L}_2}{\partial f}=0
\end{eqnarray}
and
\begin{eqnarray}\label{21}
\partial_\mu\left[\frac{\partial\mathcal{L}_2}{\partial(\partial_\mu g)}\right]  -\frac{\partial\mathcal{L}_2}{\partial g}=0
\end{eqnarray}
reduce to a single second-order equation for $f$
\begin{eqnarray}\label{22}
0
&=& \frac{1}{\lambda^2}\left(\frac{d^2f}{dr^2}+\frac{1}{r}\frac{df}{dr}-\frac{n^2}{r^2}\sin f\cos f\right) \nonumber\\
&& +~4K_s~\frac{n^2}{r^2}~\sin^2f\left(\frac{d^2f}{dr^2}-\frac{1}{r}\frac{df}{dr}\right) \nonumber\\
&& +~4K_s~\frac{n^2}{r^2}~\sin f~\cos f\left(\frac{df}{dr}\right)^2
\end{eqnarray}
with the boundary conditions $f(0)=\pi$ and $\lim_{r\rightarrow\infty}f(r)=0$ as before.

If a suitable vortex solution   of this equation existed, direct substitution into (\ref{22}) shows that it would have a series expansion for $r<<1$  of the form
\begin{eqnarray}\label{23}
f = \pi +ar+br^3+...~~\text{if}~n=1
\end{eqnarray}
or
\begin{eqnarray}\label{24}
f = \pi +ar^n+br^{3n-2}+...~~\text{if}~n\geq 2
\end{eqnarray}
where $a<0$ and $b$ are constants, and for $r>>1$ the asymptotic form
\begin{eqnarray}\label{25}
f = Ar^{-n}-\frac{1}{12}A^3r^{-3n} + ...
\end{eqnarray}
for some constant $A>0$. However, it turns out that it is not possible to match these small-distance and large-distance expansions if $K_s\neq 0$, meaning that any solution $f$ of (\ref{22}) either diverges at $r=0$ or as $r\rightarrow\infty$. This last result follows from the following simple scaling argument.

Suppose that $f(r)$ is a solution of equation (\ref{22}). Let $q$ be any positive constant and define $f_q(r)=f(qr)$. Substituting $f_q$ in place of $f$ in equation (\ref{22}) gives a value of $\mu$ which depends in general on the value of $q$
\begin{eqnarray}\label{26}
\mu_q
&=& \int\int\left\{\frac{1}{2\lambda^2}\left[\left(\frac{df_q}{dr}\right)^2+\frac{n^2}{r^2}\sin^2f_q\right] \right.\nonumber\\
&&\left. +~2K_s\frac{n^2}{r^2}\sin^2f_q\left(\frac{df_q}{dr}\right)^2\right\}r~dr~d\theta
\end{eqnarray}
where $\frac{df_q}{dr} =qf'(qr)$. So, if $r$ is replaced as the variable of integration by $\overline{r}=qr$, we have
\begin{eqnarray}\label{27}
\mu_q
&=& \int\int\left\{\frac{1}{2\lambda^2}\left[\left(\frac{df(\overline{r})}{d\overline{r}}\right)^2+\frac{n^2}{\overline{r}^2}\sin^2f(\overline{r})\right] \right.\nonumber\\
&&\left. +~2q^2K_s\frac{n^2}{\overline{r}^2}\sin^2f(\overline{r})\left(\frac{df(\overline{r})}{d\overline{r}}\right)^2\right\}\overline{r}~d\overline{r}~d\theta
\end{eqnarray}
In particular,
\begin{eqnarray}\label{28}
\frac{\partial\mu_q}{\partial q}= 4qK_s\int\int\frac{n^2}{\overline{r}^2}\sin^2f(\overline{r})\left(\frac{df(\overline{r})}{d\overline{r}}\right)^2\overline{r}~d\overline{r}~d\theta
\end{eqnarray}
But, if $f$ is a localized solution of eq.(\ref{22}), meaning that it remains suitably bounded as $r\rightarrow 0$ and as $r\rightarrow\infty$, it should (by virtue of the Euler-Lagrange equations) be a stationary point of $\mu$, meaning that $\partial\mu_q/\partial q|_{q=1}=0$. It follows therefore that no localized solution of (\ref{22}) exists.

A more rigorous statement of this property follows on from Derrick’s theorem \cite{derrick}, which states that a necessary condition for vortex stability is that
\begin{eqnarray}\label{29}
\left.\frac{\partial\mu_q}{\partial q}\right|_{q=1}=0
\end{eqnarray}
It is evident that (\ref{27}) does not satisfy this criterion.

In an attempt to fix this problem, we could add a ''mass'' term, $K_\nu(1-\hat{\textbf{z}}.\phi)$, to the Lagrangian density, $\mathcal{L}_2$, where $\hat{\textbf{z}}$ is the direction of $\phi$ at $r=\infty$ (where $f(r)=0$) and $K_\nu$ is another positive coupling constant. The Lagrangian density then becomes
\begin{eqnarray}\label{30}
\mathcal{L}_3=\mathcal{L}_2 -K_\nu(1-\hat{\textbf{z}}.\phi)
\end{eqnarray}
This Lagrangian density corresponds to the baby Skyrmion model in equation (2.2) \cite{piette}. 

The kinetic term (in the case of a free particle) together with the Skyrme term in $\mathcal{L}_2$ are not sufficient to stabilize a baby Skyrmion, as the kinetic term in $2+1$ dimensions is conformally invariant and the baby Skyrmion can always reduce its energy by inflating indefinitely. This is in contrast to the usual Skyrme model, in which the Skyrme term prohibits the collapse of the $3+1$ soliton \cite{gisiger}.
The mass term is added to limit the size of the baby Skyrmion. 

\section{Skyrmion Vortex with a Twist}
Instead of adding a mass term to stabilize the vortex, we will retain the Skyrme model Lagrangian (\ref{19}) but include a twist in the field, $g$, in (\ref{10}). That is, instead of choosing \cite{simanek},\cite{cho}
\begin{equation}\label{31}
g=n\theta-\chi
\end{equation}
we choose
\begin{equation}\label{32}
g=n\theta+mkz
\end{equation}
where $mkz$ is the twist term: as $z$ increases by an amount $2\pi$, the functions $\sin g$  and $\cos g$  oscillate a total of $mk$ times, and each oscillation corresponds to a twist in the vortex, $m$ and $n$ are integers, $2\pi/k$ is the period in the $z$-direction. 

The Lagrangian density (\ref{19}) then becomes
\begin{eqnarray}\label{33}
\mathcal{L}_2
&=& -\frac{1}{2\lambda^2}\left[\left(\frac{df}{dr}\right)^2+\sin^2f\left(\frac{n^2}{r^2}+m^2k^2\right)\right] \nonumber\\
&& -K_s\sin^2f\left(\frac{df}{dr}\right)^2\left(\frac{n^2}{r^2}+m^2k^2\right)
\end{eqnarray}

The value of the twist lies in the fact that in the far field, where $r\to\infty$ then $f\to0$, the Euler-Lagrange equations for $f$ for both $\mathcal{L}_3$ (without a twist) and $\mathcal{L}_2$ (with a twist) are formally identical to leading order, with $m^2k^2/\lambda^2$ in the twisted case playing the role of the mass coupling constant, $K_\nu$. So, it is expected that the twist term will act to stabilize the vortex just as the mass term does in $\mathcal{L}_3$.

On a physical level, the twist can be identified with a circular stress in the plane, perpendicular to the vortex string (which can be imagined e.g. as a rod aligned with the $z$-axis). The direction of the twist can be clockwise or counter-clockwise. In view of the equivalence of energy and mass, if the gravitation effects of the string were to be modelled, the energy embodied in the circular stress would contribute to the gravitational field, with the net result that the trajectories of freely-moving test particles would differ according to whether they were directed clockwise or counter-clockwise around the string.

The Euler-Lagrange equation (\ref{20}) corresponding to the twisted Skyrmion string Lagrangian density (\ref{33}) reads
\begin{eqnarray}\label{34}
0
&=&\frac{1}{\lambda^2}\left[\frac{d^2f}{dr^2}+\frac{1}{r}~\frac{df}{dr}-\frac{n^2}{r^2}~\sin f~\cos f\right] \nonumber\\
&&+~4K_s~\frac{n^2}{r^2}~\sin^2f\left(\frac{d^2f}{dr^2} -\frac{1}{r}\frac{df}{dr}\right) \nonumber\\
&& +4K_s~m^2k^2\sin^2f\left(\frac{d^2f}{dr^2}+\frac{1}{r}\frac{df}{dr}\right)\nonumber\\
&&+~4K_s\left(\frac{n^2}{r^2}+m^2k^2\right)\sin f~\cos f\left(\frac{df}{dr}\right)^2
\end{eqnarray}
It should be noted that the second Euler-Lagrange equation (\ref{21}) is satisfied identically if $g$ has the functional form (\ref{32}).

Equation (\ref{34}) can be rearranged in the form
\begin{eqnarray}\label{35}
\frac{d^2f}{dr^2}
&=&-\frac{(\varepsilon+\zeta r^2)\sin f~\cos f}{r^2+(\varepsilon+\zeta r^2)\sin^2f}\left(\frac{df}{dr}\right)^2 \nonumber\\
&& -\frac{1}{r}\left[\frac{r^2+(-\varepsilon+\zeta r^2)\sin^2f}{r^2+(\varepsilon+\zeta r^2)\sin^2f}\right]\frac{df}{dr} \nonumber\\
&&+~\frac{n^2(1+\zeta r^2/\varepsilon)\sin f\cos f}{r^2+(\varepsilon+\zeta r^2)\sin^2f}
\end{eqnarray}
where $\varepsilon=4\lambda^2K_sn^2$ and $\zeta=4\lambda^2K_sm^2k^2$ are positive constants.

Direct substitution into (\ref{35}) shows that the twisted Skyrmion string solution has a series expansion for $r<<1$ of the form
\begin{eqnarray}\label{36}
f
&=& \pi +ar^n +br^{n+2} +...
\end{eqnarray}						
where $a<0$ and $b$ are constants. The asymptotic form of the solution for $r>>1$  can be constructed by noting that, when $r$ is large and $f<<1$, equation (\ref{35}) reads
\begin{eqnarray}\label{37}
\frac{d^2f}{dr^2}\approx -\frac{1}{r}\frac{df}{dr}+\frac{n^2\zeta}{\varepsilon}f
\end{eqnarray}
This is a modified Bessel equation whose exponentially decaying solution is proportional to the modified Bessel function $K_0(n\sqrt{\zeta/\varepsilon r})$. Hence, the asymptotic form of the twisted Skyrmion string solution is
\begin{eqnarray}\label{38}
f=Ar^{-1/2}e^{-n\sqrt{\zeta/\varepsilon r}}+ ... .
\end{eqnarray}
where $A$ is a positive constant, and the remaining terms are of order $r^{-3/2}e^{-n\sqrt{\zeta/\varepsilon r}}$ or smaller.

\section{The Twisted Skyrme String Solution}
For compact twisting solutions such as the twisted baby Skyrmion string \cite{nitta1}, in addition to the topological charge, $n$, there is a second conserved quantity called the Hopf charge \cite{nitta1}. For our twisted solutions, which depend on $n\theta+mkz$, the Hopf charge is technically proportional to $nmk$. But since the topological charge $n$ is conserved, it follows that conservation of the Hopf charge is equivalent to the conservation of $mk$. The question of whether $mk$ is conserved is part of a much more general question of whether the string solutions are stable to a range of possible perturbations, both field and gravitational, but this broader question is well beyond the scope of this paper.

If we calculate the value of the Hopf charge for a twisted Skyrmion string, which is infinitely extended in the vertical direction, the charge diverges when we integrate $z$ from  $-\infty$ to $\infty$. For this reason, the Hopf charge is, strictly speaking, well defined only for compact solutions (such as in \cite{nitta1}) in which the range of $z$ is finite. Geometrically, the Hopf charge measures the number of times the solution twists through a full circle over its length in the $z$-direction. The fact that it is conserved means that if the solution is perturbed then the number of turns over its length remains fixed, no matter how it is distorted. But for twisted Skyrmion string solutions, the total number of twists is infinite because the length of the vortex is infinite, so a more useful principle in this case is that the average number of twists per unit length is conserved, or equivalently that whenever one static vortex configuration is deformed into another then $mk$ remains constant.

If $mk$ is conserved for a twisted Skyrmion string, this does not necessarily mean that the string itself is stable. There are many ways it could be unstable: it could collapse inwards to form a line singularity with infinite density, or it could expand outwards indefinitely. It is also possible that - if the self-gravity of the string is included - it might gravitationally radiate away the twists (much as a cosmic string which is almost straight but has small "bumps" is believed to radiate the energy in the bumps away). But it is typically not possible to investigate the stability of the solutions until some explicit solutions are found.

\section{Topological Charge of a Twisted Skyrmion String}
The topological charge of a general $O(3)$ field configuration is defined to be
\begin{equation}\label{39}
T=\frac{1}{4\pi}~\varepsilon_{ABC}\int\int_S \phi^A~\frac{\partial\phi^B}{\partial x}~\frac{\partial\phi^C}{\partial y}~dx~dy
\end{equation}
where $S$ is any plane parallel to the $x-y$ plane, and $\varepsilon_{ABC}$ is the Levi-Civita symbol in three dimensions, with non-zero components $\varepsilon_{123}=\varepsilon_{231}=\varepsilon_{312}=1$, $\varepsilon_{132}=\varepsilon_{213}=\varepsilon_{321}=-1$ and $\varepsilon_{ABC}=0$ if $A=B$ or $B=C$ or $C=A$.

As will be shown shortly, the topological charge $T$ is conserved, in the sense that $dT/dt=0$ for any solution (\ref{6}) of the field equations. So $T$ remains constant whenever a static vortex solution is perturbed. This charge conservation law is a consequence of the boundary
conditions  $f(0)=\pi$ and  $\lim_{r\rightarrow\infty}f(r)=0$ and does not depend on the choice of Lagrangian or equations of motion. Because the integral in (\ref{39}) is also (space) metric tensor-independent, $T$ is a purely topological quantity and for this reason is known as the topological charge \cite{mif1}.

For a twisted vortex solution constructed using the ansatz (\ref{10}), but with $n\theta-\chi$ replaced by $g(\theta,z)=n\theta+mkz$,
\begin{eqnarray}\label{40}
&&\varepsilon_{ABC}\phi^A\frac{\partial\phi^B}{\partial x}\frac{\partial\phi^C}{\partial y}\nonumber\\
&&=\varepsilon_{ABC}\phi^A\left(W^B\frac{\partial f}{\partial x}+nV^B\frac{\partial\theta}{\partial x}\right)\left(W^C\frac{\partial f}{\partial y}+nV^C\frac{\partial\theta}{\partial y}\right)\nonumber\\
\end{eqnarray}
where
\begin{equation}\label{41}
W=
\begin{pmatrix}
\cos f(r)~\sin g(\theta,z) \\
\cos f(r)~\cos g(\theta,z) \\
-\sin f(r)
\end{pmatrix};
V = 
\begin{pmatrix}
\sin f(r)~\cos g(\theta, z) \\
-\sin f(r)~\sin g(\theta, z) \\
0
\end{pmatrix}
\end{equation}
Hence
\begin{equation}\label{42}
\varepsilon_{ABC}\phi^A\frac{\partial\phi^B}{\partial x}\frac{\partial\phi^C}{\partial y}=n\varepsilon_{ABC}\phi^AW^BV^C\left(\frac{\partial f}{\partial x}\frac{\partial\theta}{\partial y}-\frac{\partial f}{\partial y}\frac{\partial\theta}{\partial x}\right)
\end{equation}
with
\begin{eqnarray}\label{43}
\varepsilon_{ABC}\phi^AW^BV^C
&=& -\sin f(r)
\end{eqnarray}
and
\begin{eqnarray}\label{44}
\frac{\partial f}{\partial x}\frac{\partial\theta}{\partial y}-\frac{\partial f}{\partial y}\frac{\partial\theta}{\partial x}=f'(r)\left(\frac{x}{r}\frac{x}{r^2}+\frac{y}{r}\frac{y}{r^2}\right)=\frac{1}{r}f'(r)
\end{eqnarray}

The topological charge (\ref{39}) is therefore
\begin{eqnarray}\label{45}
T
&=&\frac{1}{4\pi}\int_0^{2\pi}\int_0^\infty\left[-\frac{n}{r}f'(r)\sin f(r)\right]r~dr~d\theta \nonumber\\
&=& \frac{n}{2}[\cos f(r)]|_{r=0}^\infty =\frac{n}{2}[1-(-1)] = n
\end{eqnarray}
in view of the boundary conditions $f(0)=\pi$  and $\lim_{r\rightarrow\infty}f(r)=0$. The topological charge is therefore identical with the winding number $n$ of the vortex as claimed.

\section{Conservation of the Topological Charge}
We now show that the topological charge $T$ is conserved for any vortex solution, provided only that certain weak conditions are imposed on the behaviour of $\phi$ in the limit as $r\rightarrow\infty$. Taking the time derivative of (\ref{39}) gives
\begin{eqnarray}\label{46}
\frac{dT}{dt}
&=&\frac{1}{4\pi}~\varepsilon_{ABC}\int\int_S\frac{\partial\phi^A}{\partial t}~\frac{\partial\phi^B}{\partial x}~\frac{\partial\phi^C}{\partial y}~dx~dy\nonumber\\
&&+~\frac{1}{4\pi}~\varepsilon_{ABC}\int\int_S\phi^A~\frac{\partial^2\phi^B}{\partial t~\partial x}~\frac{\partial\phi^C}{\partial y}~dx~dy\nonumber\\
&&+~\frac{1}{4\pi}~\varepsilon_{ABC}\int\int_S\phi^A~\frac{\partial\phi^B}{\partial x}~\frac{\partial^2\phi^C}{\partial t~\partial y}~dx~dy
\end{eqnarray}
After integration by parts, the second integral on the right of (\ref{46}) becomes
\begin{eqnarray}\label{47}
&&\int\int_S\phi^A~\frac{\partial^2\phi^B}{\partial t~\partial x}~\frac{\partial\phi^C}{\partial y}~dx~dy\nonumber\\
&=&\int\int_S\frac{\partial\phi^B}{\partial t}\left(\frac{\partial\phi^A}{\partial x}\frac{\partial\phi^C}{\partial y}+\phi^A~\frac{\partial^2\phi^C}{\partial x\partial y}\right)~dx~dy
\end{eqnarray}
provided that $\phi$ and $\partial\phi/\partial t$ remain bounded and  $\partial\phi/\partial y\rightarrow 0$ as $r\rightarrow\infty$. Similarly
\begin{eqnarray}\label{48}
&&\int\int_S\phi^A~\frac{\partial\phi^B}{\partial x}~\frac{\partial^2\phi^C}{\partial t~\partial y}~dx~dy\nonumber\\
&&=-\int\int_S\frac{\partial\phi^C}{\partial t}~\left(\frac{\partial\phi^A}{\partial y}\frac{\partial\phi^B}{\partial x}+\phi^A~\frac{\partial^2\phi^B}{\partial x\partial y}\right)~dx~dy
\end{eqnarray}
provided that $\partial\phi/\partial x\rightarrow 0$ as $r\rightarrow\infty$.  Substitution of (\ref{48}) and (\ref{47}) into (\ref{46}) gives
\begin{eqnarray}\label{49}
\frac{dT}{dt}
&=& \frac{3}{4\pi}~\varepsilon_{ABC}\int\int_S\frac{\partial\phi^A}{\partial t}\frac{\partial\phi^B}{\partial x}\frac{\partial\phi^C}{\partial y}dx~dy
\end{eqnarray}
by virtue of the symmetries of $\varepsilon_{ABC}$.

If $\phi$ has the general form (\ref{6}) then
\begin{eqnarray}\label{50}
\frac{\partial\phi^A}{\partial t}
&=& W^A~\frac{\partial f}{\partial t}+V^A~\frac{\partial g}{\partial t};~\frac{\partial \phi^B}{\partial x}=W^B~\frac{\partial f}{\partial x}+V^B~\frac{\partial g}{\partial x};\nonumber\\
&&\frac{\partial \phi^C}{\partial y}=W^C~\frac{\partial f}{\partial y}+V^C~\frac{\partial g}{\partial y}
\end{eqnarray}
where, as before
\begin{eqnarray}\label{51}
W=
\begin{pmatrix}
\cos f~\sin g \\
\cos f~\cos g \\
-\sin f
\end{pmatrix}~\text{and}~
V = 
\begin{pmatrix}
\sin f~\cos g\\
-\sin f~\sin g \\
0
\end{pmatrix}
\end{eqnarray}
but no functional dependences are being assumed for $f$ and $g$. Given that the three vector fields  $\partial\phi/\partial t$, $\partial\phi/\partial x$ and $\partial\phi/\partial y$ in (\ref{50}) all lie in the 2-dimensional subspace spanned by $W$ and $V$, they are everywhere linearly dependent. So
\begin{eqnarray}\label{52}
\varepsilon_{ABC}\frac{\partial\phi^A}{\partial t}\frac{\partial\phi^B}{\partial x}\frac{\partial\phi^C}{\partial y}=0
\end{eqnarray}
and
\begin{eqnarray}\label{53}
\frac{dT}{dt}=0
\end{eqnarray}
as claimed.

That is, $T$ is a constant, no matter what non-linear sigma model is used.

\section{Conservation of the Hopf Charge}
Following de Vega \cite{vega} and Kobayashi and Nitta \cite{nitta1}, the Hopf charge of a twisted baby Skyrmion string (which is compact in the $z$-direction) is defined to be \footnote{The pre-factor appearing in the defining equation (9) in Kobayashi and Nitta \cite{nitta1} is $1/4\pi^2$  rather than $-1/32\pi^2$, but this is not consistent with equations (3.3)-(3.5) in de Vega \cite{vega}. See also equation (\ref{81}) below.}  
\begin{eqnarray}\label{54}
C
&=& -\frac{1}{32\pi^2}\int ~\varepsilon^{abc}~F_{ab}~A_c~dx~dy~dz
\end{eqnarray}
where $F_{ab}=\phi~.~(\partial_a\phi\times\partial_b\phi)$ is the field strength and the potential $A_c$ is a vector field satisfying the condition
\begin{eqnarray}\label{55}
F_{ab}
&=& \partial_aA_b -\partial_bA_a
\end{eqnarray}
Here, the indices $a,~b$ and $c$ range over $\{1,2,3\}$ and $\varepsilon^{abc}$  is again the Levi-Civita symbol in three dimensions, with non-zero components  $\varepsilon^{123}=\varepsilon^{231}=\varepsilon^{312}=1$ and  $\varepsilon^{132}=\varepsilon^{321}=\varepsilon^{213}=-1$. 

Like the topological charge, the Hopf charge of a twisted baby Skyrmion string is conserved provided that $\phi$ satisfies certain weak boundary conditions. This can be shown by first defining a topological 4-current
\begin{eqnarray}\label{56}
J^\mu
&=& \varepsilon^{\mu\nu\rho\sigma}F_{\nu\rho}A_\sigma
\end{eqnarray}
where $F_{\mu\nu}=\phi.(\partial_\mu\phi\times\partial_\nu\phi)$ is the full field tensor, $A_\sigma$  is a 4-potential satisfying $F_{\mu\nu}=\partial_\mu A_\nu -\partial_\nu A_\mu$, and  $\varepsilon^{\mu\nu\rho\sigma}$ is the 4-dimensional Levi-Civita symbol with $\varepsilon^{\mu\nu\rho\sigma}=1$  if  $\mu\nu\rho\sigma$ is an even permutation of 0123 and $\varepsilon^{\mu\nu\rho\sigma}=-1$  if $\mu\nu\rho\sigma$  is an odd permutation of 0123. In particular,
\begin{eqnarray}\label{57}
J^0
&=& \varepsilon^{abc}F_{ab}A_c
\end{eqnarray}

The 4-current  $J^\mu$ is conserved identically. To see this, note first that
\begin{eqnarray}\label{58}
\partial_\mu J^\mu
&=& \partial_\mu[\varepsilon^{\mu\nu\rho\sigma}(\partial_\nu A_\rho-\partial_\rho A_\nu)A_\sigma]\nonumber\\
&=& \varepsilon^{\mu\nu\rho\sigma}(\partial_\nu A_\rho-\partial_\rho A_\nu)\partial_\mu A_\sigma\nonumber\\
&=& \frac{1}{2}\varepsilon^{\mu\nu\rho\sigma}F_{\nu\rho}F_{\mu\sigma}
\end{eqnarray}
where the last expression follows because $\varepsilon^{\mu\nu\rho\sigma}=\partial_\mu A_\sigma = \frac{1}{2}\varepsilon^{\mu\nu\rho\sigma}(\partial_\mu A_\sigma -\partial_\sigma A_\mu)$.

As previously, if the $O(3)$ field $\phi$ is represented in the form (\ref{6}) then
\begin{eqnarray}\label{59}
\partial_\mu\phi = W\partial_\mu f +V\partial_\mu g
\end{eqnarray}
where W and V were defined in equation (\ref{48}). In particular,
\begin{eqnarray}\label{60}
W\times V = -
\begin{pmatrix}
\sin^2f~\sin g \\
\sin^2f~\cos g \\
\sin f~\cos f
\end{pmatrix} 
= -\phi~\sin f
\end{eqnarray}
and so
\begin{eqnarray}\label{61}
\partial_\mu\phi\times\partial_\nu\phi = -\sin f(\partial_\mu f\partial_\nu g-\partial_\nu f\partial_\mu g)\phi
\end{eqnarray}
and
\begin{eqnarray}\label{62}
F_{\mu\nu}=-\sin f(\partial_\mu f~\partial_\nu g-\partial_\nu f~\partial_\mu g)
\end{eqnarray}
as $\phi.\phi=1$.

Substituting (\ref{62}) into equation (\ref{58}) then gives
\begin{eqnarray}\label{63}
\partial_\mu J^\mu
&=& \frac{1}{2}\varepsilon^{\mu\nu\rho\sigma}F_{\nu\rho}F_{\mu\sigma}\nonumber\\
&=& \frac{1}{2}\sin^2f~\varepsilon^{\mu\nu\rho\sigma}\nonumber\\
&& \times (\partial_\nu f~\partial_\rho g-\partial_\rho f~\partial_\nu g)(\partial_\mu f~\partial_\sigma g-\partial_\sigma f~\partial_\mu g)\nonumber\\
&=&0
\end{eqnarray}
by virtue of the anti-symmetry of $\varepsilon^{\mu\nu\rho\sigma}$. The fact that $J^\mu$ is conserved allows us to write the time derivative of the Hopf charge as
\begin{eqnarray}\label{64}
\frac{dC}{dt}
&=& -\frac{1}{32\pi^2}\frac{d}{dt}\int\int\int\varepsilon^{abc}F_{ab}A_c~dx~dy~dz \nonumber\\
&=& -\frac{1}{32\pi^2}\int\int\int\partial_0J^0~dx~dy~dz\nonumber\\
&=& \frac{1}{32\pi^2}\int\int\int\partial_aJ^a~dx~dy~dz
\end{eqnarray}
The last term is an integral of the 3-divergence $\partial_aJ^a$ over some closed connected subset $\sum$ on  $\mathbb{R}^3$. So provided that the components $J^a$  are continuously differentiable on $\sum$, and the boundary  $\partial\sum$ of $\sum$ is piecewise smooth, an application of the Divergence Theorem reduces (\ref{64}) to
\begin{eqnarray}\label{65}
\frac{dC}{dt}
&=& \frac{1}{32\pi^2}\int\int_{\partial\sum}J^an_a~dS
\end{eqnarray}
where  $n_a$ is the outward unit normal on $\partial\sum$. The Hopf charge $C$ is therefore conserved if the surface integral on the right of (\ref{65}) vanishes.

In the case of the twisted baby Skyrmion string \cite{nitta1},  $\partial\sum$ can be chosen to be a cylindrical surface of constant $r$ in the compactified 3-dimensional space. Then so long as $rJ^a\partial_ar$  goes to zero as  $r\rightarrow\infty$, and $J^3$  is strictly periodic over the compactified range  $[z,z+\Delta z]$ in the $z$-direction, the Hopf charge $C$ defined by integrating over the full 3-dimensional space is conserved.

However, in the case of the twisted Skyrmion string model considered here, which exists on a non-compact 3-dimensional space, the situation is more delicate. As was mentioned in Section 5 above, the Hopf charge $C$ is typically undefined if the integral in (\ref{54}) extends over all of $\mathbb{R}^3$. It will therefore be assumed that the domain of integration in (\ref{54}) is bounded by two horizontal planes, say at $z=z_1$  and $z=z_2$, with $z_1<z_2$. The surface $\partial\sum$  can then be chosen to be a cylinder with $r$ constant on the side boundary, and  $z=z_1$ and  $z=z_2$ on the two top boundaries, and the Hopf charge $C$ will be conserved if
\begin{eqnarray}\label{66}
\int_{z_1}^{z_2}\int_0^{2\pi}rJ^a~\partial_ar~d\theta~dz\rightarrow 0
\end{eqnarray}
in the limit as $r$ goes to $\infty$, and
\begin{eqnarray}\label{67}
\int_0^{2\pi}\int_0^{\infty}J^3|_{z=z_2}~r~dr~d\theta -\int_0^{2\pi}\int_0^\infty J^3|_{z=z_1}r~dr~d\theta=0\nonumber\\
\end{eqnarray}

The distance $\Delta=z_2-z_1$ can in principle be chosen to be arbitrarily large, but the principle that the total charge $C$ between the two planes $z=z_1$ and $z=z_2$ is conserved does entail that the string vortex is in some sense "periodic" at these boundaries, as it implicitly
assumes that any twists in  $\phi$ that enter $\sum$ through one of the boundary planes is exactly balanced by the propagation of twists out of $\sum$ through the other boundary plane. A more detailed specification of the boundary behaviour required will be given at the end of the next section.

\section{Hopf Charge of a Twisted Skyrmion String}
The twisted Skyrmion string model has
\begin{eqnarray}\label{68}
\vec\phi
&=&
\begin{pmatrix}
\sin f(r)~\sin (\theta,z) \\
\sin f(r)~\cos (\theta,z) \\
\cos f(r)
\end{pmatrix}
\end{eqnarray}
where $g(\theta,z)=n\theta+mkz$, and $mkz$ is the twist term. So in view of equation (\ref{62})
\begin{eqnarray}\label{69}
F_{ab}
&=& -f'(r)\sin f(r)[(n~\partial_b\theta+mk~\partial_bz)\partial_ar \nonumber\\
&&-(n~\partial_a\theta+mk~\partial_az)\partial_br]
\end{eqnarray}
where, since
\begin{eqnarray}\label{70}
\partial_ar=\delta_a^1\frac{x}{r}+\delta_a^2\frac{y}{r};~\partial_a\theta=-\delta_a^1\frac{y}{r^2}+\delta_a^2\frac{x}{r^2};~\partial_az=\delta_a^3\nonumber\\
\end{eqnarray}
it follows that
\begin{eqnarray}\label{71}
&&(n~\partial_b\theta+mk~\partial_b z)\partial_ar \nonumber\\
&&= (n~\partial_a\theta +mk~\partial_az)\partial_br = nr^{-1}(\delta_a^1\delta_b^2-\delta_a^2\delta_b^1)\nonumber\\
&& +~mk[(\partial_ar)\delta_b^3-(\partial_br)\delta_a^3]
\end{eqnarray}
and so
\begin{eqnarray}\label{72}
F_{ab}
&=& -nr^{-1}f'\sin f(\delta_a^1\delta_b^2-\delta_a^2\delta_b^1) \nonumber\\
&& -mk~f'\sin f[(\partial_ar)\delta_b^3-(\partial_br)\delta_a^3 ] \nonumber\\
&=& -nr^{-1}f'\sin f(\delta_a^1\delta_b^2-\delta_a^2\delta_b^1) \nonumber\\
&& +mk[(\partial_a\cos f)\delta_b^3-(\partial_b\cos f)\delta_a^3]
\end{eqnarray}

A vector field $A_c$ satisfying $F_{ab}=\partial_a A_b-\partial_bA_a$ can be constructed by noting first of all that in the general case (\ref{62}) one possible choice of vector potential is
\begin{eqnarray}\label{73}
\bar{A}_\mu=\cos f~\partial_\mu g
\end{eqnarray}
as it is clear by inspection that $\partial_\mu\bar{A}_\nu-\partial_\nu\bar{A}_\mu=F_{\mu\nu}$. The most general vector potential is therefore
\begin{eqnarray}\label{74}
A_\mu=\bar{A}_\mu+\partial_\mu\omega
\end{eqnarray}
where $\partial_\mu\omega$ is an arbitrary gradient.

In the case of a twisted Skyrmion string, the last equation reads
\begin{eqnarray}\label{75}
A_c=n\cos f\left(\delta_c^2\frac{x}{r^2}-\delta_c^1\frac{y}{r^2}\right)+mk\cos f~\delta_c^3+\partial_c\lambda
\end{eqnarray}
In keeping with the symmetries of $\bar{A}_c$, the gradient term should be chosen to have the form
\begin{eqnarray}\label{76}
\partial_c\omega
&=& n\alpha~\partial_c\theta+mk\beta~\partial_cz\nonumber\\
&=& n\alpha\left(\delta_c^2\frac{x}{r^2}-\delta_c^1\frac{y}{r^2}\right)+mk\beta~\delta_c^3
\end{eqnarray}
where  $\alpha$ and  $\beta$ are constants to be determined, so that
\begin{eqnarray}\label{77}
A_c=n(\alpha+\cos f)\left(\delta_c^2\frac{x}{r^2}-\delta_c^1\frac{y}{r^2}\right)+mk(\beta+\cos f)\delta_c^3\nonumber\\
\end{eqnarray}

The constants $\alpha$ and $\beta$ are fixed by the requirement that the components of $A_c$ remain finite as  $r\rightarrow 0$, and vanish as  $r\rightarrow\infty$. Since  $f(0)=\pi$ and  $\lim_{r\rightarrow\infty}f(r)=0$, it follows
immediately that  $\alpha=1$ and  $\beta=-1$. So
\begin{eqnarray}\label{78}
A_c=n(1+\cos f)\left(\delta_c^2\frac{x}{r^2}-\delta_c^1\frac{y}{r^2}\right)+mk(-1+\cos f)\delta_c^3\nonumber\\
\end{eqnarray}
Furthermore, because $\bar{A}_\mu$ is proportional to $\partial_\mu g$, $\varepsilon^{\mu\nu\rho\sigma}F_{\nu\rho}\bar{A}_\sigma=0$ identically and so, in the general case,
\begin{eqnarray}\label{79}
J^\mu=\varepsilon^{\mu\nu\rho\sigma}F_{\nu\rho}A_\sigma=\varepsilon^{\mu\nu\rho\sigma}F_{\nu\rho}\partial_\sigma\omega
\end{eqnarray}
This means that, for a twisted Skyrmion string,
\begin{eqnarray}\label{80}
&&\varepsilon^{abc}F_{ab}A_c \nonumber\\
&&= -2f'\sin f~\varepsilon^{abc}[nr^{-1}\delta_a^1\delta_b^2+mkr^{-1}(x\delta_a^1+y\delta_a^2)\delta_b^3]\nonumber\\
&&\times \left[n\left(\delta_c^2\frac{x}{r^2}-\delta_c^1\frac{y}{r^2}\right)-mk\delta_c^3 \right] \nonumber\\
&&= 4nmkr^{-1}f'\sin f
\end{eqnarray}
and
\begin{eqnarray}\label{81}
C
&=& -\frac{nmk}{8\pi^2}\Delta z\int_0^{2\pi}\int_0^\infty(r^{-1}f'\sin f)r~dr~d\theta \nonumber\\
&=& \frac{nmk}{4\pi}\Delta z[\cos f(r)]|_{r=0}^\infty = \frac{nmk}{2\pi}\Delta z
\end{eqnarray}

Note that, in the case a twisted baby Skyrmion string \cite{nitta1}, the length of the compactified vertical dimension is  $\Delta z=2\pi$ and  $C=nmk$.

It is evident from (\ref{81}) that the Hopf charge $C$ is proportional to the distance $\Delta=z_2-z_1$ between the boundary planes. If the field  $\phi$ is perturbed away from the static twisted vortex configuration (\ref{68}), $C$ will be conserved provided that the limiting conditions (\ref{66}) and (\ref{67}) are satisfied. A general non-static field configuration can be modelled by writing
\begin{eqnarray}\label{82}
A_\mu
&=& n(1+\cos f)\left(\delta_\mu^2\frac{x}{r^2}-\delta_\mu^1\frac{y}{r^2}\right)\nonumber\\
&& +mk(-1+\cos f)\delta_\mu^3+B_\mu
\end{eqnarray}
where  $B_\mu$ represents the deviation of the vector potential from the potential of the static vortex.

For the purposes of deriving a sufficiency condition for the conservation of the Hopf charge $C$, it is not necessary to assume that the components of  $B_\mu$ are small everywhere on the subset $\sum$ over which (\ref{68}) is integrated, although it will be assumed that they tend to zero as  $r\rightarrow\infty$. In view of (\ref{72}), the components of the topological 4-current  $J^\mu$ that are relevant
to the conservation of $C$ are
\begin{eqnarray}\label{83}
&& rJ^a\partial_ar \nonumber\\
&&= -2mk(x\varepsilon^{1ab0}+y\varepsilon^{2ab0})(f'\sin f)(x\delta_a^1\delta_b^3+y\delta_a^2\delta_b^3)B_0 \nonumber\\
&& +~(x\varepsilon^{1\mu\nu\rho}+y\varepsilon^{2\mu\nu\rho})(\partial_\mu B_\nu-\partial_\nu B_\mu)B_\rho \nonumber\\
&&= (x\varepsilon^{1\mu\nu\rho}+y\varepsilon^{2\mu\nu\rho})(\partial_\mu B_\nu-\partial_\nu B_\mu)B_\rho
\end{eqnarray}
and
\begin{eqnarray}\label{84}
J^3=-2nr^{-1}(f'\sin f)B_0 +\varepsilon^{3\mu\nu\rho}(\partial_\mu B_\nu-\partial_\nu B_\mu)B_\rho \nonumber\\
\end{eqnarray}

From equation (\ref{83}), it is evident that the first Hopf charge conservation condition (\ref{66}) will be satisfied if $(\partial_\mu B_\nu-\partial_\nu B_\mu)B_\rho$ goes to zero more rapidly than $r^{-1}$ in the limit as $r\rightarrow\infty$, while from (\ref{84}) the second conservation condition (\ref{67}) will be satisfied if the integral
\begin{eqnarray}\label{85}
&& \int_0^{2\pi}\int_0^\infty[-2nr^{-1}(f'\sin f)B_0 \nonumber\\
&& +~\varepsilon^{3\mu\nu\rho}(\partial_\mu B_\nu-\partial_\nu B_\mu)B_\rho]~r~dr~d\theta 
\end{eqnarray}
has the same value on the two boundary planes $z=z_1$ and $z=z_2$ The last condition will in turn hold if the difference function $\Delta B_\mu=B_\mu|_{z=z_2} - B_\mu|_{z=z_1}$ is zero for the three components $B_0$, $B_1$ and $B_2$ and is zero also for the first derivatives of these three components with respect to $t$, $x$ and $y$.

\section{Conclusions}
We have shown that, for any non-compact $O(3)$ field configuration $\phi$, the topological charge $T$ defined by equation (\ref{39}) is conserved, provided that some weak constraints on the behaviour of $\phi$ in the limit as $r\rightarrow\infty$ are satisfied, and that in the case of the twisted Skyrmion string $T$ is equal to the winding number $n$. 

By contrast, the Hopf charge $C$ defined by equation (\ref{54}) is typically finite only if the integration region $\sum$ over which $C$ is evaluated is compact in the $z$-direction. In the case of the twisted Skyrmion string, $C$ is equal to  $(nmk/2\pi)\Delta z$, where $\Delta z$ is the vertical extent of $\sum$, and $mk$ is the twist of the vortex. The Hopf charge $C$ will also be conserved if (i) $\phi$ decays
sufficiently rapidly in the limit as $r\rightarrow\infty$ and (ii) the horizontal and time components of the field strength tensor $F_{\mu\nu}$ and vector potential  $A_\rho$ corresponding to $\phi$ are the same on the two
boundary planes $z=z_1$ and $z=z_2$.

The last condition is a relatively strong one, and requires the vortex to be in some sense "periodic" at the two boundary planes. It is clear on physical grounds that a condition of this type is needed to guarantee the conservation of Hopf charge, as in the general
unconstrained case there is nothing to stop twists in the vortex from propagating into or out of the region $\sum$. Only if the number of twists that enter $\sum$ is exactly balanced by the number of twists that leave $\sum$ would we expect the Hopf charge to be conserved, and it is this balance that condition (ii) enforces.

\begin{center}
\textbf{Acknowledgment}
\end{center}
This research was fully funded by Graduate Research Scholarship from Universiti Brunei Darussalam (GRS UBD) for MH. This support is greatly appreciated.
\\~~\\

\end{document}